\documentclass[12pt]{article}

\usepackage{amsmath,amsfonts,amssymb}

\begin{document}

\begin{titlepage}

\begin{center}

\vskip 0.4 cm

\begin{center}
{\Large{ \bf Note About Non-Relativistic Diffeomorphism Invariant
Gravity  Action in Three Dimensions}}
\end{center}

\vskip 1cm

{\large Josef Kluso\v{n}$^{}$\footnote{E-mail: {\tt
klu@physics.muni.cz}} }

\vskip 0.8cm

{\it Department of
Theoretical Physics and Astrophysics\\
Faculty of Science, Masaryk University\\
Kotl\'{a}\v{r}sk\'{a} 2, 611 37, Brno\\
Czech Republic\\
[10mm]}

\vskip 0.8cm

\end{center}

\begin{abstract}
This short note is devoted to   Hamiltonian analysis of three
dimensional  gravity action that was  proposed
recently in [arXiv:1309.7231].  We perform
this  analysis 
 and determine the number of physical degrees of
freedom.

\end{abstract}

\bigskip

\end{titlepage}

\newpage

\def\bn{\mathbf{n}}
\newcommand{\bC}{\mathbf{C}}
\newcommand{\bD}{\mathbf{D}}
\def\hf{\hat{f}}
\def\tK{\tilde{K}}
\def\bk{\mathbf{k}}
\def\tr{\mathrm{tr}\, }
\def\tmH{\tilde{\mH}}
\def\tY{\mathcal{Y}}
\def\nn{\nonumber \\}
\def\bI{\mathbf{I}}
\def\tmV{\tilde{\mV}}
\def\e{\mathrm{e}}
\def\bE{\mathbf{E}}
\def\bX{\mathbf{X}}
\def\bY{\mathbf{Y}}
\def\bR{\bar{R}}
\def\hN{\hat{N}}
\def\hK{\hat{K}}
\def\hnabla{\hat{\nabla}}
\def\hc{\hat{c}}
\def\mH{\mathcal{H}}
\def \Gi{\left(G^{-1}\right)}
\def\hZ{\hat{Z}}
\def\bz{\mathbf{z}}
\def\bK{\mathbf{K}}
\def\iD{\left(D^{-1}\right)}
\def\tmJ{\tilde{\mathcal{J}}}
\def\tr{\mathrm{Tr}}
\def\mJ{\mathcal{J}}
\def\partt{\partial_t}
\def\parts{\partial_\sigma}
\def\bG{\mathbf{G}}
\def\str{\mathrm{Str}}
\def\Pf{\mathrm{Pf}}
\def\bM{\mathbf{M}}
\def\tA{\tilde{A}}
\newcommand{\mW}{\mathcal{W}}
\def\bx{\mathbf{x}}
\def\by{\mathbf{y}}
\def \mD{\mathcal{D}}
\newcommand{\tZ}{\tilde{Z}}
\newcommand{\tW}{\tilde{W}}
\newcommand{\tmD}{\tilde{\mathcal{D}}}
\newcommand{\tN}{\tilde{N}}
\newcommand{\hC}{\hat{C}}
\newcommand{\hg}{\hat{g}}
\newcommand{\hX}{\hat{X}}
\newcommand{\bQ}{\mathbf{Q}}
\newcommand{\hd}{\hat{d}}
\newcommand{\tX}{\tilde{X}}
\newcommand{\calg}{\mathcal{G}}
\newcommand{\calgi}{\left(\calg^{-1}\right)}
\newcommand{\hsigma}{\hat{\sigma}}
\newcommand{\hx}{\hat{x}}
\newcommand{\tchi}{\tilde{\chi}}
\newcommand{\mA}{\mathcal{A}}
\newcommand{\ha}{\hat{a}}
\newcommand{\tB}{\tilde{B}}
\newcommand{\hrho}{\hat{\rho}}
\newcommand{\hh}{\hat{h}}
\newcommand{\homega}{\hat{\omega}}
\newcommand{\mK}{\mathcal{K}}
\newcommand{\hmK}{\hat{\mK}}
\newcommand{\hA}{\hat{A}}
\newcommand{\mF}{\mathcal{F}}
\newcommand{\hmF}{\hat{\mF}}
\newcommand{\hQ}{\hat{Q}}
\newcommand{\mU}{\mathcal{U}}
\newcommand{\hPhi}{\hat{\Phi}}
\newcommand{\hPi}{\hat{\Pi}}
\newcommand{\hD}{\hat{D}}
\newcommand{\hb}{\hat{b}}
\def\I{\mathbf{I}}
\def\tW{\tilde{W}}
\newcommand{\tD}{\tilde{D}}
\newcommand{\mG}{\mathcal{G}}
\def\IT{\I_{\Phi,\Phi',T}}
\def \cit{\IT^{\dag}}
\newcommand{\hk}{\hat{k}}
\def \cdt{\overline{\tilde{D}T}}
\def \dt{\tilde{D}T}
\def\bra #1{\left<#1\right|}
\def\ket #1{\left|#1\right>}
\def\mV{\mathcal{V}}
\def\Xn #1{X^{(#1)}}
\newcommand{\Xni}[2] {X^{(#1)#2}}
\newcommand{\bAn}[1] {\mathbf{A}^{(#1)}}
\def \bAi{\left(\mathbf{A}^{-1}\right)}
\newcommand{\bAni}[1]
{\left(\mathbf{A}_{(#1)}^{-1}\right)}
\def \bA{\mathbf{A}}
\newcommand{\bT}{\mathbf{T}}
\def\bmR{\bar{\mR}}
\newcommand{\mL}{\mathcal{L}}
\newcommand{\mbQ}{\mathbf{Q}}
\def\mat{\tilde{\mathbf{a}}}
\def\mtF{\tilde{\mathcal{F}}}
\def \tZ{\tilde{Z}}
\def\mtC{\tilde{C}}
\def \tY{\tilde{Y}}
\def\pb #1{\left\{#1\right\}}
\newcommand{\E}[3]{E_{(#1)#2}^{ \quad #3}}
\newcommand{\p}[1]{p_{(#1)}}
\newcommand{\hEn}[3]{\hat{E}_{(#1)#2}^{ \quad #3}}
\def\mbPhi{\mathbf{\Phi}}
\def\tg{\tilde{g}}

\section{Introduction}

 Recently
new form of three dimensional gravity action was proposed in
\cite{Andreev:2013qsa}. This action was derived
 from relativistic three dimensional gravity action by new 
limiting procedure which however shares some  similarity with the analysis performed in
\cite{Horava:2009uw} and also in \cite{Horava:2010zj}\footnote{For
review of Ho\v{r}ava-Lifshitz gravity, see
\cite{Horava:2011gd,Weinfurtner:2010hz,Sotiriou:2010wn}.}.
Among other new results it was further  argued in
\cite{Andreev:2013qsa}
  that given action
is invariant under non-relativistic diffeomorphism whose  explicit
form is given in (\ref{nonreltr}).

The form of this three dimensional gravity action
 is very interesting and certainly deserves
further investigation. 
 In particular, it would be interesting to
find the Hamiltonian formulation of given action. To do this
we rewrite the action found in \cite{Andreev:2013qsa} to more familiar
form which is similar to the three dimensional non-relativistic
action studied in \cite{Horava:2010zj}. However there is an important
difference between these actions since the action studied in 
\cite{Horava:2010zj} is invariant under foliation preserving diffeomorphism 
which is more general than 
(\ref{nonreltr}) since in addition to the rules given in 
(\ref{nonreltr}) it contains the transformation 
$t'=f(t)$. As a result the Hamiltonian formulation of the action  analyzed in
\cite{Horava:2010zj} possesses the global Hamiltonian constraint which, as 
we will see,  is absent in case of the action given 
in \cite{Andreev:2013qsa}. On the other hand the presence of the global
constraint does not restrict the number of the local degrees of freedom. 
Explicitly, when we perform the Hamiltonian analysis of the 
action given in \cite{Andreev:2013qsa} and identify all constraints  we
find that there are no local degrees of freedom with agreement with
the number of physical degrees of freedom of three dimensional 
non-relativistic covariant Ho\v{r}ava-Lifshitz gravity  \cite{Horava:2010zj}.


%

\section{ New Non-Relativistic Gravity Action} \label{second}
In \cite{Andreev:2013qsa}   new form of non-relativistic $2+1$ gravity
action was found in the form
\begin{equation}\label{Andreevact} S=m\int dt
d^2\bx \sqrt{g} \left[-\frac{1}{4}\dot{g}^{ij}\dot{g}_{ij}-
\frac{1}{4}(g^{ij}\dot{g}_{ij})^2- \frac{1}{m} (A^i\partial_i\ln g-
\dot{g}^{ij}\nabla_i A_j+A_0R)+\frac{1}{4m^2}F_{ij}F^{ij}\right] \ ,
\end{equation}
where $F_{ij}=\partial_i A_j-\partial_j A_i$.
 Note that this theory lives on a $2-$dimensional surface without
boundaries with the metric tensor $g_{ij}$, where $g=\det g_{ij}$,
$R$ is corresponding scalar curvature and where $\nabla_i$ is
covariant derivative compatible with the metric $g_{ij}$.

Before we proceed to the Hamiltonian formalism we would like to
explicitly check whether  this  action is invariant under
non-relativistic diffeomorphism
\begin{eqnarray}\label{nonreltr}
A'_0(\bx',t)&=&A_0(\bx,t)-A_i(\bx,t)\partial_t
\xi^i(\bx,t) \ ,  \nonumber \\
 A'_i(\bx',t)&=&
A_i(\bx,t)-A_j(\bx,t)\partial_i\xi^j(\bx,t)
-m g_{ij}(\bx,t)\partial_t \xi^j(\bx,t) \ , \nonumber \\
g'_{ij}(\bx',t)&=&g_{ij}(\bx,t)-\partial_i\xi^k g_{kj}(\bx,t)-
g_{ik}\partial_j\xi^k(\bx,t) \ ,  \nonumber \\
\end{eqnarray}
where
\begin{equation}
\delta t\equiv t'-t =0 \ , \quad \delta x^i\equiv x'^
i-x^i=\xi^i(t,\bx) \ .
\end{equation}
For future purposes  we rewrite the action (\ref{Andreevact})
 into  more familiar form. First of all we  replace $\dot{g}^{ij}$ with
\begin{equation}
\dot{g}^{ij}=-g^{im}\dot{g}_{mn}g^{nj} \
\end{equation}
 using the definition
of the metric $g^{ij}$ as inverse to $g_{ij}$. Then  the kinetic
term takes the form
\begin{eqnarray}
g^{ik}g^{il}\dot{g}_{ij}\dot{g}_{kl}- g^{ij}g^{kl}\dot{g}_{ij}
\dot{g}_{kl}
=\dot{g}_{ij}\mG^{ijkl}\dot{g}_{kl} \ , \nonumber \\
\end{eqnarray}
where
\begin{equation}
\mG^{ijkl}=\frac{1}{2}(g^{ik}g^{jl}+g^{il}g^{jk})-g^{ij}g^{kl} \ ,
\end{equation}
with inverse
\begin{equation}
\mG_{ijkl}= \frac{1}{2}(g_{ik}g_{jl}+g_{il}g_{jk})-g_{ij}g_{kl} \ , \quad
\mG_{ijkl}\mG^{klmn}= \frac{1}{2}(\delta_i^k
\delta_j^l+ \delta_i^l\delta_j^k) \ .
\end{equation}
Further, using  integration by parts  we observe that we can perform
following replacement in the action (\ref{Andreevact})
\begin{eqnarray}
-\int d^2\bx\sqrt{g}A^i\partial_t \partial_i\ln g= \int
d^2\bx\partial_i [\sqrt{g}A^i]\partial_t\ln g =\int
d^2\bx\partial_i[\sqrt{g}A^i]\dot{g}_{kl}g^{kl} \ .
\nonumber \\
\end{eqnarray}
Observe that  $\sqrt{g}A^i$ is tensor density of weight $1$ so that
\begin{eqnarray}
\nabla_i (\sqrt{g}A^i)=\partial_i(\sqrt{g}A^i) \ .  \nonumber \\
\end{eqnarray}
%
Now with the help of these results we  find that the action
(\ref{Andreevact}) can be rewritten into the form
\begin{eqnarray}\label{Andrejevact2}
S&=&m\int dt d^2\bx \sqrt{g}\left[\tK_{ij}\mG^{ijkl}\tK_{kl}
-\frac{1}{m}A_0R-\right.
\nonumber \\
&-&\left.\frac{1}{4m^2}(\nabla_i A_j+\nabla_j
A_i)\mG^{ijkl}(\nabla_k A_l+
\nabla_l A_k) +\frac{1}{4m^2}F_{ij}F^{ij}\right] \ , \nonumber \\
\end{eqnarray}
where we defined $\tK_{ij}$ as
\begin{equation}
\tK_{ij}=\frac{1}{2}(\partial_t g_{ij}-\frac{1}{m}\nabla_i A_j-
\frac{1}{m}\nabla_j A_i) \ .
\end{equation}
Finally we use the fact that
\begin{eqnarray}
\int d^2\bx\sqrt{g}\nabla_i A_j g^{ij}\nabla_k A_l g^{kl}=\int
d^2\bx \sqrt{g}\left(\nabla_k A_j g^{ji} \nabla_i
A_l g^{lk}+\frac{A_i A^i}{2}R\right) \ , \nonumber \\
\end{eqnarray}
where we performed integration by parts and used the definition
\begin{equation}
\nabla_i\nabla_j A_k-\nabla_j\nabla_i A_k=R_{ijk}^{ \quad l}A_l
\end{equation}
together with the fact that two dimensional Riemann tensor has the
form
\begin{equation}
R_{ijkl}=\frac{R}{2}(g_{ik}g_{jl}-g_{il}g_{kj}) \ ,
\end{equation}
where $R$ is the scalar curvature. As a result we find that the
action (\ref{Andrejevact2}) has the form
\begin{eqnarray}\label{Andrejevact3}
S=
m\int dt d^2\bx \sqrt{g}\left[\tK_{ij}\mG^{ijkl}\tK_{kl}
-\frac{1}{m}\left(A_0-\frac{A_iA^i}{2m}\right)R\right]
\ .
\nonumber \\
\end{eqnarray}
Now we explicitly show that this  action
 is invariant under (\ref{nonreltr}). To begin with note
that $\tK_{ij}$ transform under (\ref{nonreltr}) as
\begin{eqnarray}
K'_{ij}(\bx',t)=
K_{ij}(\bx,t)-\partial_i\xi^kK_{kj}(\bx,t)-
K_{ik}\partial_j\xi^k(\bx,t) \nonumber \\
\end{eqnarray}
and hence
\begin{equation}
(K_{ij}\mG^{ijkl} K_{kl})'(\bx',t) =K_{ij}(\bx,t)\mG^{ijkl}(\bx,t)
K_{kl}(\bx,t) \ .
\end{equation}
Further, using (\ref{nonreltr}) we find
\begin{eqnarray}
(A_0-\frac{1}{2m}A_i A^i)R(\bx',t')= (A_0-\frac{1}{2m}A_i
A^i)R(\bx,t) \nonumber \\
\end{eqnarray}
Then it is easy to see that the action 
(\ref{Andrejevact3})
is invariant under  (\ref{nonreltr}).

We can also derive the action 
(\ref{Andrejevact3}) 
directly  when we implement 
  the limiting procedure
 suggested in \cite{Andreev:2013qsa}
to the case of   three dimensional relativistic gravity 
action 
written in $2+1$ formalism. Explicitly 
we start with the  action
\begin{equation}
S=mc^2\int dt d^2\bx  \sqrt{-g^{(3)}}{}^{(3)}R \ ,
\end{equation}
where $m$ is mass scale and $c$ is speed of light, $g^{(3)}$ is
three dimensional metric and ${}^{(3)}R$ is corresponding curvature.
Following \cite{Andreev:2013qsa} we choose the parameterization of
the metric $g_{\mu\nu}^{(3)}$ as
\begin{equation}\label{hgans}
g^{(3)}_{\mu\nu}=\left(\begin{array}{cc} -1+\frac{2A_0}{mc^2} &
\frac{A_i}{mc} \\
\frac{A_i}{mc} & g_{ij} \\ \end{array} \right) \ .
\end{equation}
Now we  compare (\ref{hgans}) with the $2+1$ decomposition of the
metric $g^{(3)}_{\mu\nu}$
 \cite{Gourgoulhon:2007ue,Arnowitt:1962hi}
\begin{eqnarray}\label{hgdef}
{}^{(3)}g_{00}&=&-N^2+N_i g^{ij}N_j \ , \quad {}^{(3)}g_{0i}=N_i \ ,
\quad {}^{(3)}g_{ij}=g_{ij} \ ,
\nonumber \\
{}^{(3)}g^{00}&=&-\frac{1}{N^2} \ , \quad
{}^{(3)}g^{0i}=\frac{N^i}{N^2} \ , \quad
{}^{(3)}g^{ij}=g^{ij}-\frac{N^i N^j}{N^2} \ .
\nonumber \\
\end{eqnarray}
Further,  three dimensional scalar curvature in $2+1$ formalism
has the form
\begin{equation}
{}^{(3)}R= K_{ij}\mG^{ijkl}K_{kl}+R \ ,
\end{equation}
where
\begin{equation}
K_{ij}=\frac{1}{2N}\left(\frac{\partial g_{ij}}{\partial x^0}-
\nabla_i N_j-\nabla_j N_i\right) \ ,
\end{equation}
and  where $\nabla_i$ are covariant derivatives defined by the
metric $g_{ij}$ and where we ignored the boundary terms.
 When we identify (\ref{hgans}) and (\ref{hgdef}) we find
\begin{equation}\label{iden}
N_i=\frac{A_i}{mc} \ , \quad  N^2=1-\frac{2A_0}{mc^2}+ \frac{A_i
g^{ij}A_j}{m^2c^2} \ .
\end{equation}
Now we are ready to proceed with the limiting procedure when we
start with the action in $2+1$ formalism
\begin{equation}\label{act31}
S=mc^2\int d^3\bx \sqrt{g}N[K_{ij}\mG^{ijkl}K_{kl}+R] \ .
\end{equation}
Using (\ref{iden}) we find that the action (\ref{act31}) in the
limit  $c\rightarrow \infty$ takes the form
\begin{eqnarray}\label{actalt}
S&=&m\int dt d^2\bx\sqrt{g} \left[ \frac{1}{4}
\left(\dot{g}_{ij}-\frac{1}{m}\nabla_i A_j-\frac{1}{m}\nabla_j
A_i\right)\mG^{ijkl} \left(\dot{g}_{kl}-\frac{1}{m}\nabla_k
A_l-\frac{1}{m}\nabla_l
A_k\right)-\right. \nonumber \\
&-&\left. \frac{1}{m}\left(\frac{A_0}{m}-\frac{A_ig^{ij}A_j}{2m}\right)R\right] \ \nonumber \\
\end{eqnarray}
with complete agreement with the 
action (\ref{Andrejevact3}). 

Now we can proceed to the Hamiltonian analysis of (\ref{actalt}).
In fact, this analysis is  rather
straightforward and closely resembles the analysis performed in case
of non-relativistic covariant Ho\v{r}ava-Lifshitz gravity
\cite{Horava:2010zj,Kluson:2011xx,Kluson:2010zn}. Explicitly, from
(\ref{actalt}) we find following momenta
\begin{equation}
\pi^{ij}=\frac{\delta L}{\delta \partial_t
g_{ij}}=2m\sqrt{g}\mG^{ijkl}\tK_{kl} \ ,  \quad  \pi^i=\frac{\delta
L}{\delta
\partial_t A_i}\approx 0 \ , \quad  \pi^0=\frac{\delta L}{\delta \partial_t
A_0}\approx 0 \ .
\end{equation}
Then it is easy to see that the Hamiltonian  has the form
\begin{equation}
H=\int d^2\bx (\mH_0+\frac{1}{m}A_ig^{ij}\mH_j+v_0\pi^0+v_i\pi^i) \ ,
\end{equation}
where
\begin{eqnarray}
\mH_0&=& \frac{1}{m\sqrt{g}} \pi^{ij}\mG_{ijkl}\pi^{kl}+
\sqrt{g}\left(\frac{A_0}{m}-\frac{A_ig^{ij}A_j}{2m}\right)R \ ,
 \nonumber \\
   \mH_i&=&-2g_{ik}\nabla_j \pi^{kj} \ ,
\end{eqnarray}
and where the primary constraints $\pi^i\approx 0, \pi^0\approx 0$
were  included.

Now the requirement of the preservation of the primary constraints
$\pi^0\approx 0 \ , \pi^i\approx 0$ implies following secondary
constraints
\begin{eqnarray}
\partial_t\pi^0&=&\pb{\pi^0,H}=
-\frac{\sqrt{g}}{m}R\equiv -\frac{1}{m}\mG^0\approx 0 \ , \nonumber \\
\partial_t\pi^i&=&\pb{\pi^i,H}=
-\frac{1}{m}g^{ij}\mH_j+\frac{\sqrt{g}}{m}g^{ij}A_jR\approx
-\frac{1}{m}g^{ij}\mH_j\approx 0 \ . \nonumber \\
\end{eqnarray}
Then the total Hamiltonian with all constraints included has the
form
\begin{eqnarray}
H_T=\int d^2\bx (\mH_0+v_0\pi^0+v_i\pi^i+ \Gamma_0\mG^0+
\Gamma^i\mH_i) \ ,  \nonumber \\
\end{eqnarray}
where $v_0,v_i,\Gamma_0,\Gamma^i$ are Lagrange multipliers
corresponding to the constraints $\pi^0,\pi^i,\mG^0,\mH_i$.

As the final step we  check the stability of all constraints.  First of
all the constraints $\pi^0,\pi^i$ are  trivially preserved that
implies that they are the first class constraints.   Further we
 introduce the smeared form of the constraint $\mH_i$
\begin{equation}
\bT_S(N^i)=\int d^2\bx N^i\mH_i \ .
\end{equation}
Then it is easy to show that $\bT_S(N^i)$ is generator of the
spatial diffeomorphism with following Poisson brackets
\begin{eqnarray}\label{pbbtS}
\pb{\bT_S(N^i),\bT_S(M^j)}&=&
\bT_S(N^j\partial_jM^i-M^j\partial_jN^i)
\ , \nonumber \\
\pb{\bT_S(N^i),\mG^0}&=&-N^i\partial_i \mG^0-\partial_iN^i\mG^0 \ .
\nonumber \\
\end{eqnarray}
This result also implies  that the requirement of the preservation
of the constraints $\mH_i$ during the time evolution of the system
does not generate any additional constraint. Finally the time
development of the  constraint $\mG^0$ is given following equation
\begin{eqnarray}\label{partmG0}
\partial_t\mG^0&=&\pb{\mG^0,H_T}=
-\frac{1}{m}\nabla_m g^{mn}\mH_n \approx 0 \nonumber \\
\end{eqnarray}
using
\begin{equation}
\delta R=\nabla^i\nabla^j\delta g_{ij}-g^{ij}\nabla^k\nabla_k\delta
g_{ij}-R^{ij}\delta g_{ij} \ , \quad R_{ij}=\frac{R}{2}g_{ij} \ .
\end{equation}
Equation  (\ref{partmG0}) tells us that $\mG^0$ is preserved during
the time evolution. Further, since it weakly  commutes with all
other constraints it is the first class constraint.

Let us conclude our results. We have following collections of the
first class constraints $\pi^i\approx 0 \ , \pi^0\approx
0,\mH_i\approx 0$ and $\mG^0\approx 0$. The constraints
$\pi^i\approx 0,\pi^0\approx 0$ can be gauge fixed so that $A_i,A_0$
and their conjugate momenta are eliminated. At the same way by gauge
fixing $\mH_i,\mG^0$ we eliminate all degrees of freedom
corresponding to three dimensional metric $g_{ij}$. In other words
there are no physical degrees of freedom left. This result is in agreement with 
the number of physical degrees of freedom in 
three dimensional non-relativistic covariant Ho\v{r}ava-Lifshitz
gravity
\cite{Horava:2011gd}.
\\
 \noindent {\bf
Acknowledgement:}

I would like to thank M. Haack for his  very important remark that
leaded to the correction of the first version of this paper.
 This work   was
supported by the Grant agency of the Czech republic under the grant
P201/12/G028. \vskip 5mm

\end{document}